\documentclass[showpacs,
aps,floatfix]{revtex4}
\usepackage{amsmath,amssymb,eucal,graphicx,subfigure,hyperref}
\usepackage{color}
\begin{document}

\title{Atomic Torsional Modal Analysis for high-resolution proteins}

\author{Monique M. Tirion}
\email{mmtirion@clarkson.edu}

\author{Daniel ben-Avraham}

\affiliation{Physics Department, Clarkson University, Potsdam, New York
13699-5820, USA} 

\begin{abstract}
We introduce a formulation for normal mode analyses of globular proteins that significantly improves
on an earlier, 1-parameter formulation \cite{tirion96} that characterized the slow modes
associated with protein data bank structures.  Here we develop that empirical potential function 
which is minimized at the outset to include two features essential to reproduce the
eigenspectra and associated density of states in the $0$ -- $300\,{\rm cm}^{-1}$ frequency range, not merely the slow modes.
First, introduction of preferred dihedral-angle configurations via  use of torsional
stiffness constants eliminates anomalous dispersion characteristics due to insufficiently
bound surface sidechains, and helps fix the spectrum thin tail frequencies ($100$ -- $300\,{\rm cm}^{-1}$). Second, 
we take into account the atomic identities and the distance of separation of all pairwise interactions, improving the spectrum distribution in the $20$ -- $300\,{\rm cm}^{-1}$ range.
With these modifications not only does the spectrum reproduce that of full atomic potentials, but we obtain stable, reliable eigenmodes for the slow modes and over a wide range of frequencies.

\end{abstract}

\maketitle

%
Proteins, nanomachines of the living world, provide a bewildering array of inter and intra-cellular services: 
catalysis, transport, scaffolding, support, cell-signaling and replication among others.  
Each protein functions with extraordinary almost paradoxical efficiency, specificity and accuracy. 
A single molecule of acetylcholine esterase, for example, breaks down 25,000 neurotransmitters molecules
per second, a rate that seems to be diffusion-limited.

Proteins are linear polymers of 20 different amino acid building blocks whose sequence specifies all covalent bonds.
The so-called primary sequence of over 8 million proteins are known, and vary in length from a few dozen to
tens of thousands of amino acids.
The primary sequences fold into distinct secondary patterns of helices, sheets and loops that
in turn combine to create distinct globular arrangements that range in cross-sectional diameter from ten
to hundreds of Angstroms.  X-ray crystallographers have determined the atomic arrangements
of nearly 100,000 protein molecules, revealing recurring folding patterns and structural motifs \cite{PDB}.

Proteins are not rigid molecules. Ever since the first crystal structures, of hemoglobin and myoglobin, revealed the
absence of entry and exit channels for ligand to reach the buried heme binding site \cite{perutz},
motility was understood to be vital to function \cite{frauenfelder}.
Protein molecules possess motility spectra over time-scales of at least 14 orders of magnitude,
from the femtosecond regime for bond breaking, to the pico- and nanosecond regimes for events like
flipping of interior and surface indole and benzene rings, to the micro- to millisecond
regimes for conformational changes such as occur during allosteric re-arrangements of secondary
and tertiary structural elements, to millisecond to seconds for rigid-like tumbling. 
As each of these dynamic processes is best characterized by different physical theories such as
quantum vs. classical mechanics, diffusion and Brownian motion, 
characterizing and sorting the many dynamic features of proteins is challenging.

Ingenious methods to track local and global motions experimentally continue to be developed using
X-rays, NMR spectroscopy, and inelastic and quasi-elastic neutron scattering, to
name a few \cite{kern}.   
Experiments detect complex net motility spectra of proteins in particular environments,
and isolating specific molecular events from the data remains difficult.
Theory segregates events and models catalytic,
diffusive and allosteric events separately into fast local and relatively slower nonlocal motions.

Rapid, local catalytic events are well-studied and modeled using quantum mechanical and
detailed semi-empirical formulations \cite{warshel}.  Extended, global motions are less well
characterized as they are hard to isolate experimentally and require lengthy simulations.
One attractive candidate to characterize global motion is normal mode analysis (NMA)
where instantaneous, equilibrium motility spectra are computed \cite{go,karplus,levitt85}.
The highly idealized condition of an absence of unbalanced forces permits  rapid and
simple snapshots of the internal flexibilities of proteins, unattainable by standard MD, and provides
fascinating insights into protein architecture and design.

Do proteins ever exhibit the motility spectra suggested by NMA?
Experimental studies, from Mossbauer spectroscopy, inelastic neutron scattering and
X-ray crystallography, demonstrate the existence of equilibrium or harmonic motion in proteins only
below a critical ``glass transition" temperature of around 180-220K \cite{petsko}.  
Do the eigenmodes then retain relevance at physiological temperatures?
At least two lines of evidence suggest they do.  High resolution X-ray crystallography provide isotropic
and sometimes anisotropic Debye-Waller values for each scattering center. These experimental values, 
a reflection of the net mobility of each atom in the crystalline environment, are surprisingly well-modeled
by the normal mode spectrum of an isolated molecule.  Second, and equally surprising perhaps,
is the successful use of normal modes to model non-equilibrium, allosteric transitions between
known, crystal structures of apo- and holo-forms of the same enzyme~\cite{petrone,bray,omori,wolynes}.  
Both lines of inquiry suggest that the very simple and idealized normal mode computation 
can provide helpful insights above the glass transition temperature as well.

Normal mode analyses provide different levels of detail depending on the formulation adopted
in setting up the algorithm. The use of atomic coordinates vs.~reduced coordinates, internal 
Cartesian vs.~dihedral  coordinates, simple vs.~detailed energy potentials, as
well as the resolution and stability of the coordinate file all affect the
computed eigenfrequency spectra and associated amplitudes as well as the appearance of each
independent mode, over the entire frequency range.  

Since demonstrating that an idealized, one-parameter energy potential reproduces very well the overall appearance of
temperature factors and therefore the slow modes that contribute  primarily to
this distribution~\cite{tirion96}, the question remained: What details were sacrificed? 
Might it be useful to more faithfully reproduce eigenfrequency spectra observed by spectroscopy
and standard  NMAs that use full, molecular mechanic potentials and 
exact molecular topology
\cite{na,song}?
Is it indeed possible, within the Hookean approximation of
\cite{tirion96}, to achieve the level of detail seen in standard NMA? If yes, we could (a)~begin to
correlate observed spectral characteristics with specific structural features, and (b)~obtain more realistic flexibility descriptions of proteins whose structures are known.

The current work represents an effort to address these questions.
We demonstrate that use of (1)~complete atomic coordinate files, (2)~dihedral coordinates,
and (3)~more realistic energy formulations than single-parameter Hookean springs 
can more realistically reproduce  experimental and theoretical vibration spectra.
Dubbed ATMAN for Atomic Torsional Modal ANalysis, this method provides a rapid visual
assessment of the internal flexibility characteristics inherent in the overall design of
each protein and may be useful to study design characteristics that contribute
to observed maxima in spectroscopic data. 

The general formalism for NMA was presented in \cite{levitt85,go,karplus}.  
Briefly, given a protein consisting of $N$ atoms of mass $m_l$, located at $\mathbf{r}_l$, 
and a potential energy function $E_p(\mathbf{r})$, one selects a proper set of generalized degrees of freedom $\{q_i\}_{i=1}^M$ and solves the co-diagonalization problem
\begin{equation}
\mathbf {FA = \Lambda H A},
\end{equation}
where the elements of the  {\em force matrix} $\mathbf F$ (also known as the {\em Hessian}),  and the mass-weighted {\em inertia matrix}
$\mathbf H$,  are:
\begin{equation}
\label{FandH.eq}
F_{ij} = \left.\frac{\partial^2E_p}{\partial q_i \partial q_j}\right|_{q=0},\qquad
H_{ij} = \sum_{l=1}^{N} m_l \frac{\partial \mathbf{r}_l}{\partial q_i}\cdot
                                             \frac{\partial \mathbf{r}_l}{\partial q_j},
\end{equation}
and the potential is assumed to be minimized at $q_i=0$.
The eigenfrequencies are given by the elements of the diagonal matrix $\mathbf{\Lambda}$; $\omega_i^2=\Lambda_{ii}$,
and the corresponding eigenmodes are the columns of $\mathbf A$, 
normalized according to $\mathbf {A^\dagger HA} = \mathbf I$. With this normalization, the amplitude $\alpha_i$ of mode $i$, resulting from thermal activation at temperature $T$, is 
$\alpha_i=\sqrt{k_BT}/\omega_i$.  Thus, the dominant contribution to the total mean-square fluctuation, given by
the sum over all the modes, comes from the low-frequency modes.

The various NMA approaches differ mainly in their choice of the degrees of freedom $\{q\}$ and the
form (and level of detail) of the potential energy function, $E_p$.  Especially for proteins whose structure files
are incomplete 
or known to low resolution,  a suitable approach is to select the Cartesian coordinates of reduced masses centered at the $C_{\alpha}$ carbons, and select an $E_p$ suitable for such a coarse-grained description \cite{eyal,phillips}. 
For high resolution structures, coarse-graining of the atomic coordinates is unnecessary.   

We strongly advocate the use of dihedral degrees of freedom to compute slow modes of proteins.
When comparing two protein sequences, one in an active, folded form and one in an unfolded, extended form,
very little variation in bond lengths and bond angles exists. The two forms vary mostly in the 
distribution of 4-atom rotational torsional degrees of freedom
(for this reason, algorithms to predict protein folding using dihedral angles abound \cite{singh}).
We use the ``soft" torsional angles of proteins: all
main-chain (MC) $\phi$ and $\psi$ as well as side-chain (SC) $\chi$ angles (except for the MC
$\phi$ angles of prolines) as detailed in the L79 potential of Levitt \cite{L79}. Not only does this reduce the
size of the computation compared to $3N$ atomic coordinates, but more importantly use of torsional degrees of
freedom necessarily maintains stereochemically correct bond lengths and bond angles throughout.

A suitable potential energy $E_p$ is needed to quantify the energetic costs of distorting the
 protein's configuration along each internal coordinate $q_i$.  Detailed
semi-empirical formulations such as  ENCAD~\cite{ENCAD} and CHARMM \cite{CHARMM} 
exist that accurately model the various types of  relevant interactions: van der Waals, hydrogen bonds, etc.
A drawback of these potentials for the current use is that the protein configuration as read from the PDB,  $\{\mathbf{r}_i^0\}$, is not initially at a local minimum, as required by a NMA.  Minimizing a complex potential function of hundreds of parameters is problematic: The potential landscape consists of many close local minima, making it difficult to compare among the results from different studies; and the minima one finds numerically exhibit various unstable modes, due to inaccuracies, casting doubt on the frequencies and shape of the remaining,
stable modes.

In previous work~\cite{tirion96} we proposed that a  potential {\em centered around the {\em PDB} configuration} of the protein gives a reasonable description of the modes.  Specifically, we suggested  
\begin{equation}
E_p = \sum_{(a,b)} E(\mathbf{r}_a,\mathbf{r}_b),
\end{equation}
where the sum is restricted to non-bonded atom pairs separated by at least three bond lengths and a certain interaction range, $R_{ab}$ (see below).
Each pair-wise potential term followed a simple Hookean form:
\begin{equation}
\label{C.eq}
E(\mathbf{r}_a,\mathbf{r}_b) = \frac{C}{2} ( \left| \mathbf{r}_{ab} \right| - 
\left| \mathbf{r}^0_{ab}\right|)^2,
\end{equation}
where $\mathbf{r}_{ab}=\mathbf{r}_b-\mathbf{r}_a$, and $|\mathbf{r}^0_{ab}|$ is the initial separation between
each interacting $a,b\,$-pair as read from the PDB file.  This potential has the obvious advantage of being at a well defined (stable) minimum at the outset.

In~\cite{tirion96} we also argued that for
the {\em slow} modes of densely packed  globular proteins one can use a single universal value for the 
spring constants $C$, since they depend on a large number of non-covalent interactions (NBI) whose sum approaches a universal form, governed by the central limit theorem, regardless of the details of individual pair-wise potentials. 
The actual value of the universal spring constant was adjusted to fit other NMA work.
We now discuss how to generalize this single bond-strength formulation, 
and replace $C$ with atom- and distance-dependent non-bonded
stiffness constants, $C_{ab}$, as well as add a new stiffness constant, $K$, for the dihedral angles.  
The advantages, as we will see,
include more realistic frequency spectra and a greater range of applicability due to the greater stability
of side chains that may be unbound or minimally bound in the PDB coordinate file.

The general principle is as follows.  Given the PDB coordinates of a protein, $\{\mathbf{r}^0\}$, and a reliable empirical potential function $E_p(\{\mathbf{r}\})$, we assume that the minimum of $E_p$ lies nearby 
$\{\mathbf{r}^0\}$~\cite{summa}.  In that case, one can approximate the Hessian $\mathbf{F}$, from Eq.~(\ref{FandH.eq}), by simply evaluating the second derivatives at $\{\mathbf{r}^0\}$ instead of at the minimum of $E_p$.  The fact that first derivatives are 
{\em not} zero at $\{\mathbf{r}^0\}$ is not a problem because they do not contribute to the Hessian.  This can be seen by Taylor-expanding pair-wise potentials around  the PDB coordinates:
$V_{ab}(r_{ab})=V_{ab}(r_{ab}^0)+V_{ab}'(r_{ab}^0)(r_{ab}-r_{ab}^0)+(1/2)V_{ab}''(r_{ab}^0)(r_{ab}-r_{ab}^0)^2+\cdots$.  The contribution of  $V_{ab}'$  to $\mathbf{F}$ vanishes if the relation between the ${\mathbf r}$ coordinates and the generalized degrees of freedom is linear, $r_{ab}-r_{ab}^0=\sum_ic_iq_i$, as coded in the $\mathbf{H}$ matrix.
Since we are replacing the potential with second-order terms, ignoring first derivatives, this guarantees an {\em extremum} point.  To ensure that the point is a {\em minimum} and the PDB configuration represents a stable
equilibrium we exclude the domain where $V_{ab}''<0$.

\begin{figure}[h]
\includegraphics[width=0.5\textwidth]{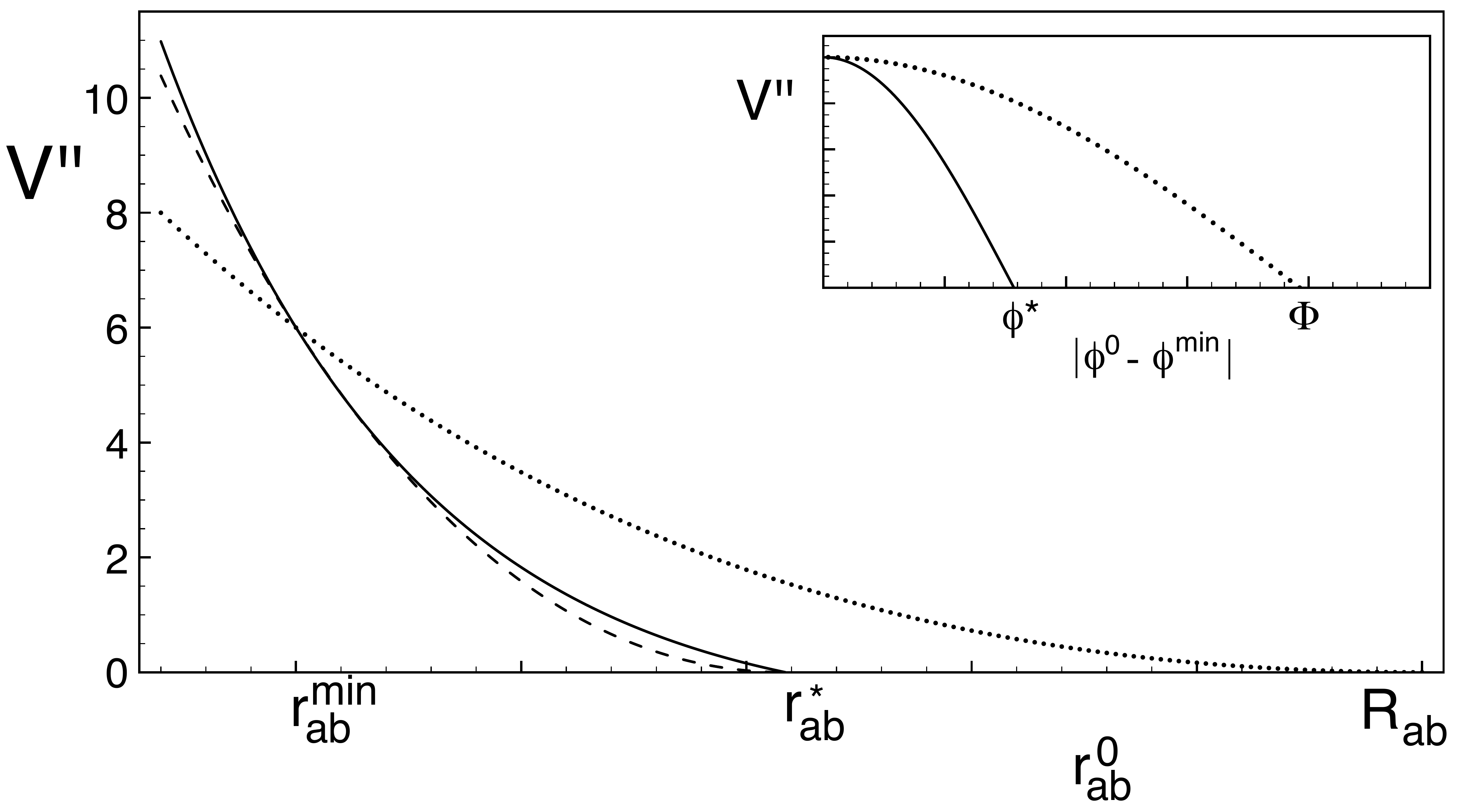}
\caption{Potential energy curvature and ``stretching": Plotted is $V''_{ab}$ as a function of $r^0_{ab}$ as given by Eq.~(\ref{V''.eq}) (solid curve), as approximated by~(\ref{V''approx.eq}) (dashed curve), and after stretching of $r_{ab}^*$ to
$R_{ab}$, Eq.~(\ref{V''effective.eq}) (dotted curve).  Both $V''_{ab}$ and $r^0_{ab}$ are shown in arbitrary units
(same units for all three cases). Inset: Stretching of the cutoff coordinate $\phi^*$ to $\Phi$ for the dihedral potential, see Eqs.~(\ref{Vd''.eq}) 
and~(\ref{Vdapprox".eq}).}
\label{v''.fig}
\end{figure}

Consider, for example, a van der Waals (vdW) force between atoms $a$ and $b$ given by the 
Lennard-Jones 6-12 potential 
\begin{equation}
V_{ab}=\epsilon_{ab}\left[\left(\frac{r^{\rm min}_{ab}}{r_{ab}}\right)^{12}
-2\left(\frac{r^{\rm min}_{ab}}{r_{ab}}\right)^{6}\right],
\end{equation}
where $\epsilon_{ab}=\sqrt{\epsilon_a \epsilon_b}$ and $r_{ab}^{\rm min}$ are the potential well-depths and the distance between $a$ and $b$ at the minimum of $V_{ab}$, respectively, while
$r_{ab}$ is the distance between the atoms.  Note that $r_{ab}^{\rm min}$ need not necessarily equal the starting value from the PDB file, $r_{ab}^0$. For $V_{ab}$'s contribution to the Hessian, it suffices to consider only the second-order derivative:
\begin{equation}
\label{V''.eq}
V_{ab}''(r_{ab}^0)=\frac{12\epsilon_{ab}}{(r_{ab}^0)^2}\left[13\left(\frac{r^{\rm min}_{ab}}{r_{ab}^0}\right)^{12}
-7\left(\frac{r^{\rm min}_{ab}}{r_{ab}^0}\right)^{6}\right].
\end{equation}
This expression turns negative for $r_{ab}^0>(13/7)^{1/6}r_{ab}^{\rm min}\equiv r_{ab}^*$, and we must exclude that domain in order to avoid potential instabilities.  
Chopping off the domain results, however, in an effectively reduced interaction range between the $a,b$ atoms.  In practice, empirical potentials  consider interactions between atoms $a$ and $b$ up to a certain range $R_{ab}$ that is computed by adding an arbitrary cutoff distance $r_c$ to the sum (or a multiple of the sum) of the van der Waals radii:
\begin{equation}
\label{Rab.eq}
R_{ab}=(13/7)^{1/6}r_{ab}^{\rm min}+r_c,
\end{equation}
and we would like to extend our interaction range all the way to $R_{ab}$ as well.  This can be done by ``stretching"
our domain.  We first note that $V_{ab}''$ admits a simpler (but approximate) form:
\begin{equation}
\label{V''approx.eq}
V_{ab}''(r_{ab}^0)\approx\frac{72\epsilon_{ab}}{(r_{ab}^0)^2}\frac{(r_{ab}^0-r_{ab}^*)^2}{(r_{ab}^{\rm min}-r_{ab}^*)^2}\,,
\end{equation}
see Fig~\ref{v''.fig}.

The stretching of the interaction range is then achieved by mapping $r_{ab}^*\mapsto R_{ab}$, to yield
\begin{equation}
\label{V''effective.eq}
V_{ab}''(r_{ab}^0)\approx \frac{72\epsilon_{ab}}{(r_{ab}^0)^2}\frac{(r_{ab}^0-R_{ab})^2}{(r_{ab}^{\rm min}-R_{ab})^2}\,.
\end{equation}
This atom- and distance-dependent spring strength
$V''_{ab}(r_{ab}^0)$ replaces the universal constant $C$ of Eq.~(\ref{C.eq}).

As a second example consider the potential function for the rotation of the dihedral angles coordinates.  
Typically:
\begin{equation}
V_{\rm dihedral}(\phi)=\frac{V_0}{2}[1-\cos n(\phi - \phi^{\rm min})],
\end{equation}
where $V_0$ is the potential depth, $n$ is the number of minima along the rotation coordinate $\phi$, and $\phi^{\rm min}$ is the value of $\phi$ at the potential's minimum.  Once again, if we are using dihedral angles as our 
internal coordinates, only the second derivative contributes (in this case, to the $F_{\phi\phi}$ entry):
\begin{equation}
\label{Vd''.eq}
V''(\phi^0)=\frac{n^2V_0}{2}\cos n(\phi^0-\phi^{\rm min}).
\end{equation}
To ensure stability, we  limit ourselves to the domain $|\phi^0-\phi^{\rm min}|<\pi/2n\equiv\phi^*$ where $V''>0$.
The reduced $\phi^0-\phi^{\rm min}$ interaction range that results from this procedure can be stretched from $\phi^*$ to a larger arbitrary cutoff $\Phi$:
\begin{equation}
\label{Vdapprox".eq}
V''(\phi^0)=\frac{n^2V_0}{2}\cos \frac{\pi(\phi^0-\phi^{\rm min})}{2\Phi}.
\end{equation}
Each of the torsional degrees of freedom can then be endowed with a potential  $V(\phi)=\frac{1}{2}V''(\phi^0)(\phi-\phi^0)^2$, where $V''(\phi^0)$ is computed as above, with the appropriate values of $V_0$, $\phi^{\rm min}$ and $n$ prescribed by the empirical parent potential.
In practice, for our purposes of fitting the total distribution of modes to that of a detailed potential, we find that it is sufficient to assume a universal constant $V''(\phi^0)=K$ for all torsional degrees of freedom (instead of the more detailed expression~(\ref{Vdapprox".eq})) and to take the range of $\phi$ to include the whole circle.

It should be clear from these examples that any potential formulation, whatever its complexity and degree of sophistication, can be treated
similarly.  We note that our procedure of cutting off the range of the variables when 2nd derivatives become negative (unstable) is not actually necessary in a properly minimized configuration.  Without this procedure,
our analysis leads to instabilities, and the cutting of the range (and extending and stretching to some additional
arbitrary distance, to compensate) is the price one pays for approximating the real minimum by the PDB configuration.

\begin{figure}[h]
\includegraphics[width=0.95\textwidth]{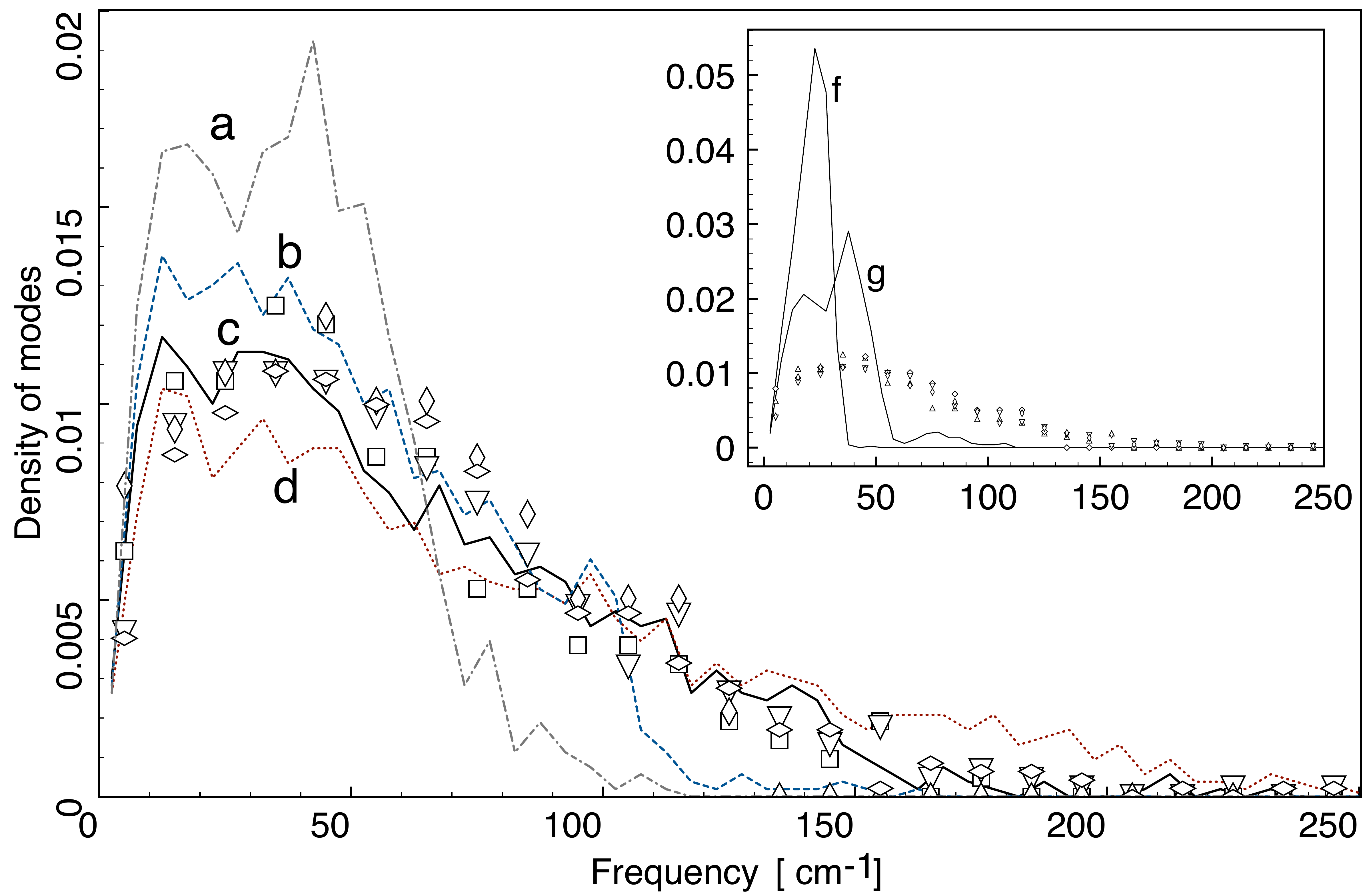}
\caption{(Color online.) Histogram of the number of modes with frequencies in $5cm^{-1}$ intervals as a function
of frequency. The curves (a)-(d) represent results obtained with decreasing cutoff values of $r_c=2.4\AA$, $1.8\AA$,
$1.6\AA$ and $1.5\AA$. The inset shows the distributions obtained using the earlier formulation from
\cite{tirion96} (f), as well as that formulation combined with dihedral stiffness constants (g).
Data obtained for 4 energy-minimized structures using a standard energy formulation
are from Levitt \cite{levitt85}, and include data for crambin, trypsin inhibitor, ribonuclease and
lysozyme,  shown as open symbols.}
\label{gw.fig}
\end{figure}

While our formulation of the potential energy is still oversimplified, it
retains the advantage of defining the initial configuration to be an exact
energy minimum, obviating the need for structure-distorting energy minimizations and the concomitant
risk of negative eigenvalues due to incomplete minimizations.  The simple potential of~\cite{tirion96}  
yielded slow modes (below the $20\,{\rm cm}^{-1}$ range) that were comparable to those found using 
more sophisticated, detailed potentials.  However, for $\omega>20\,{\rm cm}^{-1}$ the distribution of the modes deviated markedly from that obtained by standard methods (see curve (f) in the inset of Fig.~\ref{gw.fig}).
In the following we show that making the single, non-bonded stiffness strength $C$ from \cite{tirion96} atom-type and distance dependent, as in
Eq.~(\ref{V''effective.eq}), plus adding a simple dihedral potential with the same constant $K$ for the
various torsional degrees of freedom,  achieves a better description of
the {distribution of modes} throughout the whole range of frequencies (up to $300\,{\rm cm}^{-1}$).  Interestingly,
a recent independent study, focusing on B-factors rather than on the distribution of modes, by H.~Na and G.~Song~\cite{na} reached a very similar conclusion: Including detailed van der Waal interactions and a universal dihedral potential achieves the best fit to standard NMA in their excellent and comprehensive case study.

\begin{figure}[h]
\includegraphics[width=0.5\textwidth]{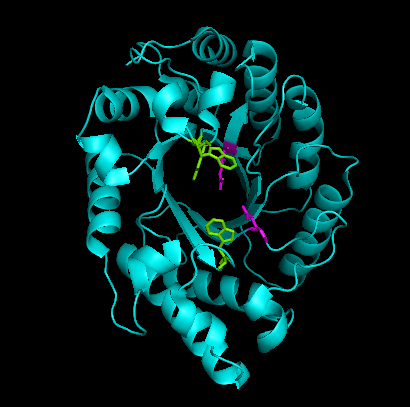}
\caption{(Color online.)  Xylanase 10A (PDB file 1GOK~\cite{leggio}) in a ribbon representation. The molecule is oriented
so as to look down the central $\beta$-barrel tube with the $\alpha$-helices arrayed on the
outside.  The resultant ``bowl''-like structure is viewed from the top with the larger cross-section.
This orientation, maintained in the left-hand panels of the GIF animations in~\cite{gifs}, shows the
active site, chemical ``scissors", Glu-131 and Glu-237, in magenta (dark gray), and the conserved
tryptophan triad, Trp-87, Trp-267 and Trp-275, in green (light gray).}
\label{ribbon.fig}
\end{figure}

To demonstrate, we computed the normal modes of xylanase 10A (PDB entry 1GOK 
confomer A), a 302 amino-acid
enzyme with a mass of nearly 33,000 daltons produced by a thermophilic ascomycetous fungus.
Xylanase 10A, a member of the family 10 glycoside hydrolases, cleaves internal bonds of xylan, 
a major polysaccharide component of plant cell walls,  releasing shorter xylose subunits.
The crystal structure of 1GOK has been determined to $1.14\AA$ resolution \cite{leggio} and is seen to
possess a classic TIM barrel fold, Fig.~\ref{ribbon.fig}, consisting of eight parallel $\beta$-strands forming a central tube
connected by 8 $\alpha$-helices external to the $\beta$-strand tube. The 
structure has been likened to a `salad bowl' as one face
has a larger radius, of about $45\AA$, while the opposite face has a smaller radius of
about $30\AA$~\cite{collins}. The active site consists of a substrate-binding cleft running the length of the larger 
face of the `salad bowl' and includes two conserved residues, Glu-131 and Glu-237, 
that act as the catalytic proton donor and nucleophile for the hydrolysis. The crystal structure
reveals that three conserved tryptophans, at locations 87, 267 and 275, 
form an aromatic cage around the active site.

To compute the normal modes associated with this structure, we built in MC amide hydrogens \cite{reduce}
to obtain a 2598-atom system with 589 MC $\psi$  and $\phi$ and 471 SC $\chi$ dihedrals,
resulting in a total of 1060 degrees of freedom.  We computed the {\bf F} and {\bf H} matrices in
Eq.~(\ref{FandH.eq}) using both analytic and numeric algorithms to check for consistency.   The use of
non-Cartesian degrees of freedom necessitates elimination of overall translations and rotations
about the center of mass in the computation of  $\mathbf{H}$.
Failure to do so results in incorrect frequencies and eigenmodes, including zero- or numerically unstable modes.

Our original formulation of the Hookean NMA  \cite{tirion96} set the dihedral stiffness constants to zero.
Compared to the cumulative non-bonded interactions of thousands of atom pairs, the contribution
to the total energy of each dihedral was considered negligible.   Careful examination of the
eigenspectra, however, revealed that the slowest modes often included ones resulting in 
root mean square (RMS) displacements discontinuous with neighboring modes. Studying the animations of these
anomalous eigenmodes revealed isolated surface sidechains sweeping wide arcs, independent of
the remainder of the molecule.  These side chains were characterized by relatively small numbers of
interatomic interactions to stabilize their conformations, which resulted in the anomalous RMS
displacements.  In past studies, we replaced such residues with alanine residues lacking sidechain degrees
of freedom, obviating the problem. However, introducing physically plausible constraints on the
dihedral bond deformations removes the need to alter the PDB coordinates. 
We found that a $K$ value of $0.1\,{\rm kcal}/{\rm mol}/ {\rm rad}^2$ stabilized all surface
side chains and resulted in reliable and accurate eigenspectra. 

For the non-bonded, vdW forces we used Eq.~(\ref{V''effective.eq}) with 
$r_{ab}^{min}=2^{1/6}[R_{vdW}(a)+R_{vdW}(b)]/2$,
to identify all non-bonded atom pairs separated by at least 3 bond lengths and within an interaction distance  
$R_{ab}$ given by Eq.~(\ref{Rab.eq}). With $r_c$ 
values ranging from 1.5 to 2.4 $\AA$, this resulted in 8,000 to 21,000 non-bonded interactions.
The resulting stiffness constants were computed using Eq.~(\ref{V''effective.eq}) and set to
$C_{ab} = CV^{''}_{ab}(r^0_{ab})$.  $C$ is a constant introduced to adjust the overall scale of the eigenfrequencies and was chosen, in each case, to make the frequency of mode 1 equal to  
$2\,{\rm cm}^{-1}$.  This required $C$ values ranging from 160. to 4.0.
The values of the van der Waals radii, $R_{vdW}$, and the Lennard-Jones well depths $\epsilon$ were
taken from ENCAD~\cite{ENCAD}.

With the various van der Waals radii and $\epsilon_{ab}$ determined from standard potentials, there remain only three parameters to adjust: The cutoff distance $r_c$; the overall
constant $C$, in the relation $C_{ab }= C V''_{ab}$; and the dihedral curvature constant $K$.
The distribution of modes of globular proteins (at all frequencies), as obtained from standard potentials, tend all to fall under one universal curve~\cite{tirion93,universality} and we use this curve to fit our parameters.  The universal
curve provides two
important clues, as $K$ and and $C$ have very different effects on the shape of the distribution.
This, coupled with the fact that the slowest mode of such a protein  as xylanase is expected to be around 
$2\,{\rm cm}^{-1}$
constrains the value of the three parameters  quite effectively~\cite{remark}.

The resulting eigenfrequency spectra, $g({\omega})$, are presented in Fig.~\ref{gw.fig} as histograms with $5\,{\rm cm}^{-1}$ bin size.
The spectra~(a-d) for different numbers of NBI
are superposed on the eigenvalue spectra of various proteins computed
by Levitt \cite{levitt85} using a standard energy formulation and energy-minimized coordinates.  In an effort to
better characterize the effects of the current formulation, we include in the inset of Fig.~\ref{gw.fig}
equivalent spectra of 1GOK using the earlier formulation presented in \cite{tirion96}(f), as well as that formulation
combined with the $K=0.1$ constraints on the dihedrals~(g).
In all cases, the low-frequency modes overlap, indicating that the various formulations identify the character
of  the slowest modes equally well.  The simplest Hookean formulation~(f), using identical spring-constants for all
non-bonded interactions and no dihedral constraints, obtains a sharp peak and narrow distribution centered around 
$25\,{\rm cm}^{-1}$, with no high-frequency tail. Interestingly, inelastic neutron scattering experiments by
Cusack and Doster identify such a well-defined maximum in the dynamic structure factors at around $25\,{\rm cm}^{-1}$
in myoglobin for temperatures below 180K \cite{cusack}. This peak is well resolved in all NMA formulations we tested.

Inclusion of dihedral stiffness constraints adds a second maximum to these spectra, as seen in (g), around $45\,{\rm cm}^{-1}$,
widening the distribution and introducing higher-frequency, tail-end contributions. While higher frequency
contributions are observed, the additional peak at
 $45\,{\rm cm}^{-1}$ is not reported on in experiments and not resolved by standard NMA,  and indicates that
the formulation remains overly simplistic. 
Use of scaled non-bonded stiffness constants adjusted the spectra significantly (a-d).
Decreasing the number of NBI has the effect of widening the distribution over a
larger frequency range and de-emphasizing the $45\,{\rm cm}^{-1}$ peak. The largest number of NBI
tested, at 21,000 [curve~(a)], produces a distribution similar to that of (g) that used a single
stiffness constant, $C$ for all interacting atom pairs.  At $r_c$ equal to 1.6\AA, resulting in
9000 nonbonded interactions, we obtained the best fit to the theoretical curve of Levitt.
Decreasing the cutoff further tends to spread the distribution even more evenly, as seen in (d) for
$r_c=1.5\AA$.
 
Studying each of these cases (a-e) via animations as well as RMS fluctuations
 demonstrates an interesting pattern, a pattern further clarified in the limit of zero non-bonded
interactions.  In the case of zero NBI and only dihedral stiffness constraints, we obtain
an eigenspectrum with purely one-dimensional traits: a broad distribution over the entire frequency range,
and a density of states $g(\omega)\sim\omega^0$ for small frequencies.
As the number of NBI
increases, the spectrum begins to shift toward that of a two-dimensional solid, with distinct peaks
and with $g(\omega)\sim\omega^1$. In particular,
for cutoff values that
permit fewer than about 5,000 non-bonded interactions, the spectra remain largely two-dimensional and
no ``hitching" of neighboring (non-bonded) regions toward cohesive motion is observed. As the number of
NBI increase, neighboring strands begin to ``hitch" together and move cohesively.
Increasing the NBI still further strengthens the cohesion and larger domains move in lockstep.
The largest cutoff values lead to a motility spectrum exhibiting very smooth
motions more typical of three-dimensional solids with $g(\omega)\sim\omega^2$.

Larger cutoffs resulting in more cohesive motions appeal to the eye in animations due to their ultra-smooth,
rubber-like deformations.  However, the computed frequency spectra indicate that this regime does
not best model either spectroscopic or theoretical results, which indicate a broader maximum and
softer molecule, even in the case of the temperature regime below 180K.  We find that small cutoff
distances leading to minimal cohesion, or ``hitching,'' best reproduce eigenspectra characteristics.
This condition also ensures sensible activation
amplitudes relevant to computing mean square displacement values and describe stereochemically sound structural deformations.  (What constitutes ``minimal cohesion" seems to depend on the resolution of structures, with
lower-resolution structures obtaining larger cutoff values than high-resolution structures.)

Animated GIF sequences of the first three slowest modes are available at~\cite{gifs}.
These sequences were created by PyMol and rendered in GIF format by LICEcap, and use
a surface rendering mode to simplify the representation of the many atoms.
In each left-hand panel, the structure is oriented as in Fig.~\ref{ribbon.fig}
so as to look at the top of the ``salad bowl" where the ligand-binding groove and active site are situated.
The active site residues, Glu-131 and Glu-237, are colored magenta, while the active-site cage formed
by the triad Trp-87, Trp-267 andTrp-275, is colored green.  The ligand-binding groove runs horizontally,
with subsites binding nonhydrolyzed subunits  binding to the left of the active site
and subsites holding hydrolyzed reactants to the right. In mode 1, the molecule was rotated around
a vertical axis by $90^\circ$ to create the right-hand panel.  In mode 2, the molecule was rotated
around a vertical axis to create the view in the right-hand panel looking down from the top of
the molecule to demonstrate the twisting nature of the motion. In mode 3, the molecule was
rotated around a vertical axis to see the bottom of the molecule in the right-hand panel.

What is immediately apparent is that while the
computation of eigenmodes proceeds without any information about ligand-binding propensities
and active site residues information, the slowest 3 modes each center on precisely these regions.  It would
seem that structure has been selected to promote dynamic access to the active site and
ligand-binding groove.  The slowest mode pertains to an up-and-down chewing type motion, the second
slowest mode resembles a rotational grinding motion, while the third slowest mode describes
a side-wise rolling type motion.  These eigenmodes appear ideally suited to enhance entry of ligand
and release of products from the binding groove, and to possibly influence the translocation of
the polysaccharide chain along the binding-groove.  While detailed, mechanistic models may be unwarranted
by these simple calculations and visualizations, it would appear that protein architecture co-evolved with
catalytic functionality, precisely in order to exploit innate flexibility characteristics to optimize yield.

To better appreciate the combination and distribution of dihedral angles contributing to each mode and
their associated domain motions, we overwrote the Debye-Waller temperature factor, $B$, in the PDB
coordinate file with normalized and binned dihedral values for each eigenmode vector. 
To visualize the  distribution of the dihedral strengths for each mode, we associated
all the $\phi$ torsions with their associated MC amide nitrogens and all the $\psi$ torsions to
the MC carbonyl carbons.

Studying the distribution of dihedral shifts for mode 1, for example, explains how the C-terminal and N-terminal
domains move {\it en-masse}, and how the overall appearance is of two domains that come together
across the ligand-binding groove, as in the first animation~\cite{gifs}. 
$\psi$ angles in mode 1 with normalized values of
0.5 or higher include
Gly-41, Pro-45 and Pro-200.  $\phi$ angles with values over 0.5 include
Gly-41, Ile-201 and Thr-208. These ``hinge-points" permit the relative reorientation of regions, with
the N and C terminal regions moving as a unit above the binding-groove.
Gly-41 at the N-terminal end of $\beta$-strand 8 is seen to move in phase with the upper domain, while
Pro-45 at the C-terminal end of the same strand moves in phase with the lower domain. 
Likewise, Pro-200 preceding $\beta$-strand 4 moves
in phase with the lower domain, while Thr-208 at the C-terminal end of the same strand moves
along with the upper domain.  So mode 1 would seem to have an upper domain extending
across residues 1-45 as well as residues 208-302, while the lower domain seems to extend from
residues 46-200.  

In this manner we tried to isolate  ``hot spots" that enable this motion and sought to mutate
these hinge-points to suppress the observed motility pattern.  Suspecting Gly-41, Pro-45,
Pro-200 and  Thr-208 as essential to the motion inherent in mode 1, we ``mutated'' these residues
by eliminating their associated MC dihedral degrees of freedom (reducing the
size of the computation by 8 degrees of freedom, in other words).  Suspecting we might
suppress mode 1 in this way, we wondered if mode 2 would now make the dominant contribution
to the RMS fluctuations. In fact, this did not happen. Mode 1 reappeared as before, but with a
different distribution of hinges, to once again present as two domains that come together
across the binding-groove.  And this seems to be a general finding: the slow modes are a
characteristic of the overall architecture of the molecule, independent of any particular residues
(publication in preparation).

In conclusion, we find that NMA does support efforts to characterize global protein 
motion on relatively slow time scales.
A straightforward computation that characterizes the full motility spectrum at a single instant, 
NMAs require neither excessive memory nor CPU capabilities (current computations were run on
a Mac Mini with a single Intel Core i5 processor).  Visualizations of eigenmodes via
animations, using the many tools available for rendering atomic coordinate files, are
effortless.  A simple sequence of 5 to 9 PDB entries, flashed consecutively on the screen,
creates a dynamic effect with a level of realism not attainable using reduced coordinates.

Furthermore, features of the eigenfrequency spectra contain additional clues to structural characteristics.
Cusack and Doster, for example, deduced that the experimentally observed maximum at 
around $25\,{\rm cm}^{-1}$ is not due to internal deformations
of secondary structures, but rather to cross-chain interactions, such as the relative
motions of $\alpha$ helices~\cite{cusack}.  Our current analyses support this interpretation, as
a reduction in the non-bonded cutoff range eliminates the ``knitting together" of non-bonded regions 
and the pronounced peak at $25\,{\rm cm}^{-1}$.
Furthermore, we find the appearance of a second peak, near $45\,{\rm cm}^{-1}$, by the inclusion
of dihedral stiffness constants.  The height and width of this additional peak
depends sensitively on the value(s) of $K$ chosen, as does the appearance of the resultant high-frequency
tail-end of the distribution. Here we used a small value, equal
to $0.1\,{\rm kcal}/{\rm mol}/{\rm rad}^2$ across all dihedrals, simply in order to stabilize the structure.
However, it is certainly possible to adopt more realistic values for each dihedral angle in order to
more faithfully reproduce the spectra.  

Inclusion of dihedral stiffness constants stabilizes surface side-chain residues and C- and N-terminal
residues that typically obtain less packing constraints than inner residues, and enable  more
reliable diagonalizations, ones that do not obtain anomalous slow frequencies.  However, these spectra
still lack the width typical of standard NMA. In order to reproduce this characteristic, we needed to
soften or lessen the energetic contribution of nonbonded atom pairs at greater distances. Cutting
the nonbonded cutoff distance to the smallest permissible while still maintaining minimal
cohesion between nonbonded strands is helpful in softening the distribution but does not appear
to be sufficient.  We find that de-emphasizing the contributions of pairwise interactions at larger
distances recreates the soft, wide distribution successfully.  Dividing the total numbers of 
NBI by the number of atoms in the molecule provides a crude estimate of the numbers
of nonbonded interaction per atom, on average.   Values over 12/atom lead to stiff molecules with narrow distributions
and  steeper slopes at small frequencies.  
Optimal results occur when the average numbers of interactions per atom are
no more than 6, which would seem to include only nearest neighbor interactions, within a single
``shell".  Interestingly, at these smaller cutoff values, a shift in the distribution of types of nonbonded
interactions occur. If we divide the list of non-bonded interactions into 3 groups, for those that
pertain to MC-MC, MC-SC and SC-SC interactions, the largest group depends on the cutoff value.
In particular, at small cutoff values, MC-MC type interactions predominate, while as the cutoff increases,
MC-SC type interactions become dominant, perhaps providing further rationalization of how 
linear chains folded into compact shapes obtain spectra with spectral dimension greater than 1 but
less than~2~\cite{elber}.

How might solvent affect computed modes?  Frequency shifts due to solvent depend on whether
proteins are overdamped or underdamped in their aqueous environments.  Oscillators in underdamped
environments experience no shifts in their eigenfrequencies while overdamped oscillators do.
Experimental evidence favors the underdamped situation for proteins. Recently, for example,
Turton and collaborators reported results with extremely sensitive optical Kerr-effect spectroscopy
of lysozyme \cite{turton}.  
They found global vibrational modes that are underdamped, and concluded
that ``the ligand-binding coordinate in proteins is underdamped and not simply solvent-controlled
as previously assumed'' \cite{turton}.  The current analysis adds to this intriguing finding by
demonstrating motility patterns centered on the active site.

We conclude that slow (ATMAN) modes of PDB entries can provide novel insights to crystallographers and 
biochemists who
aim to elucidate structure and function correlates.  While ``a picture is worth a thousand words,"  
we suspect  an animation is worth another hundred or so and foresee a future where each PDB
entry may obtain another tab for visualizing the consequence of its particular architecture by
gifs demonstrating the first few slowest modes, and with the possibility of downloading such
sequence files to study that information independently \cite{wako,song}.

\acknowledgements

MMT gratefully acknowledges support from Michael Levitt who emailed every two or three years
to assure us that (a)~modes were still very much in vogue and (b)~were really easy to do nowadays.
Without his continued support my interest would not have been reawakened, and I cannot
sufficiently express my gratitude for this.  MMT also thanks M. Hildred Blewett whose generous
bequest provided the means to re-activate this research. 
This work is supported in part by the M. Hildred Blewett Fellowship of the American Physical
Society, \url{http://www.aps.org}.

%


\begin{thebibliography}{99}
\bibitem{remark} This last requirement --- the
frequency of the slowest mode --- can be replaced by a more robust statistic, e.g., the fact that about 14\%
of the modes fall below $20\,{\rm cm}^{-1}$.
\end{thebibliography}
\end{document}